\documentclass[12pt]{article}
\usepackage{amssymb}
\usepackage{epsfig}
\usepackage[english]{babel}
\usepackage{float}
\usepackage{slashed}
\newcommand{\be}{\begin{equation}}
\newcommand{\ee}{\end{equation}}
\newcommand{\bea}{\begin{eqnarray}}
\newcommand{\eea}{\end{eqnarray}}

\def\p{\partial}
\def\pslash{\p\raise.3ex \hbox{\kern-.5em /}}
\def\delslash{\nabla\raise.3ex \hbox{\kern-.7em /}}

\begin{document}
\vskip 5cm

\begin{center}
\Large{ \textbf{\cal Non-Hermitian $\cal PT$-symmetric quantum
mechanics of relativistic particles with the restriction
 of mass}}
\end{center}
\vskip 0.5cm\begin{center} \Large{V.N.Rodionov}
\end{center}
\vskip 0.5cm
\begin{center}
{RGTU, Moscow, Russia,  \em E-mail    vnrodionov@mtu-net.ru}
\end{center}

\begin{center}

\abstract{ The modified Dirac equations for the massive particles
with the replacement of the physical mass $m$  with the help of
the
relation  $m\rightarrow m_1+ \gamma_5 m_2$  are investigated.
It is shown that for a fermion theory with a $\gamma_5$-mass term,
the limiting of the mass specter  by the value $ m_{max}=
{m_1}^2/2m_2$ takes place. In this case the different regions of
the unbroken $\cal PT$ symmetry may be expressed by means of the
restriction  of the physical mass $m\leq m_{max}$. It should be
noted that in the approach which was developed by C.Bender et al.
for the $\cal PT$-symmetric version of the massive Thirring model
with $\gamma_5$-mass term, the region of the unbroken $\cal
PT$-symmetry was found in the form $m_1\geq m_2$ \cite{ft12}.
However on the basis of the mass limitation $m\leq m_{max}$ we
obtain  that  the domain $m_1\geq m_2$ consists of two different
parametric sectors:

 i) $0\leq m_2 \leq m_1/\sqrt{2}$ -this  values of mass
parameters $m_1,m_2$  correspond to the traditional particles for
which in the limit $m_{max}\rightarrow \infty$ the modified models
are converting  to the ordinary Dirac theory with the physical
mass $m$;

ii)$m_1/\sqrt{2}\leq m_2 \leq m_1$ -   this is the case of the
unusual particles for  which equations of motion does not have a
limit, when $m_{max}\rightarrow \infty$. The presence of this
possibility lets hope for that in Nature indeed there are some
"exotic fermion fields".

 As a matter of fact the formulated criterions  may be
 used as  a major test in the process of the division of
 considered models into
 ordinary and exotic fermion theories. It is tempting
to think that the quanta of the exotic fermion field have a
relation to the structure of the "dark matter".     }

\end{center}


\section{Introductory remarks}

As it is well-known the idea about existence of a maximal mass in
a broad spectrum of elementary particles  was suggested by
M.A.Markov \cite{Markov}.  After  that  a more radical approach
was developed by V.G. Kadyshevsky and his colleagues.   Their
model  contained a limiting mass $M$ as a new fundamental phisical
constant. Doing this condition of finiteness  of the mass spectrum
should be introduced by the relation: \be\label{Mfund1} m \leq
M.\ee Really in the papers \cite{Kad1}-\cite{Rod1} the existence
of mass $M$ has been understood as a new principle of Nature
similar to the relativistic and quantum postulates, which was put
into the grounds of the new quantum field  theory. At the same
time the new constant $M$ is introduced  in a purely geometric
way, like the velocity of light is the maximal velocity in the
special relativity.

However the studying of  questions of discrete symmetries of this
theory were completely disregarded.  That is why  many long years
nobody takes no notice the existence of non-Hermitian Hamiltonians
which arise in the fermion sector of the theory with the maximal
mass. Indeed, if one chooses a geometrical formulation of the
quantum field theory, the adequate realization of the limiting
mass hypothesis  is reduced to the choice of the de Sitter
geometry as the geometry of the 4-momentum  space of the constant
curvature with a radius equal to $M$.

For example, introducing the designation of $ p_\mu
=i\partial_\mu$ and taking into account, on the mass shell
$p_5=\sqrt{M^2-m^2}$
 we have the analogue of the Dirac equation \cite{Max},\cite{KMRS}:

 \be \label{DM}\Big(p_0 - {\bf\hat \alpha p}- \hat\beta m_1 -
\hat\beta\gamma^5 m_2\Big )\Psi(x,t,x_5=0)=0 \ee were

\be\label{dm1} m_1=2M\sin\mu/2, m_2=2M\sin^2\mu/2 , \sin\mu =
m/M.\ee

In the modified Dirac equation matrix $\hat\beta =\gamma_0$,
$\gamma^i=\hat\beta\hat\alpha^i$. It is important to note that on
the mass shell $p_5=M\cos\mu$  there are not operators, which act
on the coordinate of $x_5$, and this parameter without loss of
generality can be set equal to zero \cite{Max},\cite{KMRS}.

We see that the Hamiltonian which are associated with equations
(\ref{DM}) can be represented in the form \be\label{H1} {\hat{H}}
= \overrightarrow{\hat{\alpha}}\overrightarrow{p} + \hat{\beta
}\left(m_1 + m_2\gamma_5\right). \ee It is easily checked that in
a flat limit $M\rightarrow \infty $ the (\ref{DM}),(\ref{H1}) goes
into the standard Dirac expressions. It is obvious that the
expression (\ref{H1}) is non-Hermitian due to the appearance in
them of $\gamma_5$-mass components ($H\neq H^{+}$).

A.Mustafazadeh identified the necessary and sufficient
requirements of reality  of eigenvalues for pseudo-Hermitian and
$\cal PT$-symmetric Hamiltonians and formalized the use these
Hamiltonians in his papers \cite{alir} \cite{ali}. According to
\cite{alir} and \cite{ali} we can define  Hermitian operator
$\eta$, which transform non-Hermitian  Hamiltonian \be\label{H+}
H^+ =\alpha p+ \beta(m_1 -\gamma_5 m_2)\ee
 by means of invertible
transformation to the Hermitian-conjugated one \be \eta H
\eta^{-1}= H^{+},\ee
 where $\eta = e^{\gamma_5 \alpha} $; $\alpha=
\arctan (m_2/m_1)$ \cite{Rod1}.

Now it is well-known fact the reality of the spectrum is a
consequence of $\cal PT$ -invariance of the theory, i.e. a
combination of spatial and temporary parity of the total
Hamiltonian: $[H,PT]\psi =0$.  When the $\cal PT$ symmetry is
unbroken, the spectrum of the quantum theory is real.  These
results explain the growing interest in this problem. For the past
a few years studied a lot of new non-Hermitian $\cal PT$-invariant
systems (see, for example, \cite{alir}- \cite{most5}).  In the
literature, which was devoted to the study not Hermitian operators
there are examples, with the $\gamma_5$ mass extension.

 In particular the modified Dirac equations for
the massive Thirring model in two-dimensional space-time
 with the replacement of the physical mass $m$ by $m\rightarrow m_1+ \gamma_5 m_2$
 was investigated by Bender et al.\cite{ft12}. The region of the unbroken
$\cal PT$-symmetry has been found in the form \cite{ft12}

\be\label{Thir} m_1\geq m_2 \ee .

However, it is not apparent that the area with undisturbed $\cal
PT$-symmetry $ m_1 \geq m_2 $ does not  include also the sectors,
corresponding to the  some unusual particles, description of which
radically distinguish from traditional one. These been  observed
in the paper \cite{Rod1} as "exotic particles" (see also
\cite{Max},\cite{KMRS}). Consequently the question arises: "what
precisely particles are considered: exotic or traditional, when
the condition (\ref{Thir}) is executed?"

Indeed according to the analysis, carried out in the work
\cite{Rod1}, the expressions (\ref{DM}),(\ref{H1}) can  also be
rewritten  as \be\label{A2} \Big(p_0 - {\bf\hat \alpha p}-
\hat\beta m_3 - \hat\beta\gamma^5 m_4\Big )\Psi(x,t,x_5=0)=0 \ee
 and \be\label{H2}
{\hat{H}} = \overrightarrow{\hat{\alpha}}\overrightarrow{p} +
\hat{\beta }\left(m_3 + m_4\gamma_5\right). \ee
where\footnote{Note that similar designations for the masses
(\ref{dm1}),(\ref{dm2}) has been used also in the works
\cite{neznamov1}, \cite{neznamov2}.}

\be\label{dm2} m_3=2M\cos\mu/2, m_4=2M\cos^2\mu/2\ee

 The distinguishing feature of
expressions (\ref{A2}),(\ref{H2}) consists of the fact that they
have not the limit when $ M\rightarrow \infty $, i.e. there are
not
 values of parameters in order to  obtain the ordinary Dirac
expressions. Thus, one can assume that in this case we deal with a
description of some new particles, properties of which have not
yet been studied.

This paper has the following structure. In section II the
necessary and sufficient conditions are formulated  for the case
of the restriction of the mass spectrum of particles in considered
models. In the third section we study the basic characteristics of
$\cal PT$-symmetry of free fermion models with $\gamma_5$ a
massive contribution and show that the area of unbroken $\cal PT$
symmetry in which the mass spectrum is real have an internal
structure.

\section{Necessary and sufficient conditions  of the mass
spectrum restrictions  in the model with $\gamma_5$ mass term. }

Consider now the algebraic approach developed in the numerous
papers on the study of quantum non-Hermitian mechanics.
 It was possible to expect, that the appearance of the models
described by Hamiltonians  type (\ref{H1}) is the prerogative of a
purely geometric approach to the construction of a modified theory
with a maximal mass. However, in the paper
 \cite{ft12} was considered the ${\cal PT}$-symmetric
{massive Thirring model } in (1+1)-dimensional space. As the
foundation of this study  is assumed the a model with the density
of the Hamiltonian, which is represented in the form:

\be\label{B1} {\cal
H}(x,t)=\bar{\psi}(x,t)\Big(-i\overrightarrow{\partial}\overrightarrow{\gamma}+m_1+\gamma_5
m_2 \Big)\psi(x,t). \ee The equation of motion associated with the
(\ref{B1}), may be expressed  as

\be\label{B2} \Big(i\partial_\mu\gamma^{\mu}-m_1-\gamma_5 m_2
\Big) \psi(x,t)=0, \ee that on the form coincides with the
equation (\ref{DM}).

In this regard, there is interest in the study of the parameters,
characterizing of the masses, included in the $\gamma_5$ - theory.
Note that the $\gamma_5$ - extension of the mass in the Dirac
equation consists of replacing $m\rightarrow m_1+\gamma_5 m_2$ and
after that two new mass parameters: $m_1$ and $m_2$ arises. When
the Dirac equation converts  to the Klein-Gordon equation: \be
\label{KG} \left(\partial^2+m^2\right)\psi(x,t)=0 \label{e20} \ee
there is a relation \be \label{012 } m^2={m_1}^2- {m_2}^2. \ee It
is easy to see that the physical mass $m$, appearing in the
equation (\ref{KG}), is real when the inequality \be \label{e210}
{m_1}^2\geq {m_2}^2.\ee is accomplished.

The algebraic formalism developed in \cite{ft12}, contains no
indications of the existence of other restrictions, in which
participates the physical mass of particles $m$, in addition to
(\ref{012 }). However, on the basis of  the coincidence of
equations (\ref{DM}) and (\ref{B2}), we can assume that in this
model restriction of the type inequality (\ref{Mfund1})
\textbf{should also be present.} It is very important fact, that
this conditions should be obtained because with its help may be
established the connection between algebraic and geometric
approaches. Note also that the existence of such restrictions, in
particular, may essentially modify  the fundamental results
obtained in the paper \cite{ft12}.

Let us consider after \cite{ft12} the relativistic quantum
mechanics with the $\gamma_5$-mass term in the case of 1+1
dimensional space-time. We introduce the following 2d
representation of a $\gamma$-matrices \cite{2D}: \be
\gamma_0=\left(\begin{array}{cc}0 & 1\\ 1 & 0
\end{array}\right) \quad \quad
\gamma_1=\left(\begin{array}{cc}0 & 1\\ -1 & 0 \end{array}\right).
\label{e14} \ee According to these definitions, $\gamma_0^2=1$ and
$\gamma_1^2=-1$. We also have the \be \gamma_5=
-\gamma_0\gamma_1=\left(\begin{array}{cc}1&0\\0&-1\end{array}\right),
\label{e15} \ee so that $\gamma_5^2=1$.

 Consider the Hamiltonians of the type (\ref{H1}) and
show that the the algebraic approach allows one to set the
\textbf{sufficient condition} of the limitations of the mass
spectrum of particles in relativistic quantum mechanics with
$\gamma_5$-mass component. As noted above, the inequality
(\ref{e210}) was considered in the paper \cite{ft12} as  the
single requirement that determines the presence  broken or
unbroken  $\cal PT$ - symmetry of the Hamiltonian. However, it is
easy to show that (\ref{e210}) may not be as the single condition.

Writing the following obvious inequality: \be (m-m_2)^2 \geq 0 \ee
and taking into account (\ref{012 }), we obtain \be \label{M}m
\leq \frac{{m_1}^2}{2 m_2}= m_{max},\ee that is direct indication
of the existence of the  restriction of the physical mass $m$ in
the considered model.

It is interesting that (\ref{M}) is obtained as the result of the
simple algebraic transformation of relationships with subsidiary
parameters of mass $m_1$ and $m_2$. It is quite natural that the
value of the $m_{max}$ is expressed through a combination of them.
In particular, as the degree of deviation of the Hamiltonian $H$
of a Hermitian forms is characterized by the mass  $m_2$, then its
value can be expressed through the $m_1,\,\,m_{max}$ and, taking
advantage of the symbol $$ \frac{m_1}{2 m_{max}}=\sin \theta\leq
1,$$ rewrite (\ref{H1}),(\ref{H2}) in the resulting form
\be\label{eps} H= \vec{\hat{\alpha}}\vec{p}+\hat{\beta}
m_1\Big(1+\gamma_5 \sin \theta\Big), \ee where the area
$0\leq\theta < \pi/4 $ - corresponds to the theory  having a "flat
limit" and $\pi/4 <\theta\leq \pi/2 $ - refers to the occasion
when it is absent. The first condition corresponds to the
description of ordinary particles and the second - to unusual or
exotic particles. The limit value $\theta = \pi/4 $ is responsible
for the particles with the maximal mass, which was named by the
maximons \cite{Markov}.

Thus, the limitation of the mass spectrum of particles (\ref{M}),
described by Hamiltonians (\ref{H1}),(\ref{H2}) and (\ref{eps} is
a simple consequence of the $\gamma_5$-mass extension in
(\ref{DM}),(\ref{A2}). Therefor the presence of the non-Hermitian
Hamiltonian (\ref{eps}),  in essence, can be interpret as  the
\textbf{sufficient condition} of the limitation of the mass
spectrum of particles in fermion models with $\gamma_5$ - massive
term.

On the other hand, we have  that the introduction of the new
physical postulate, which is connected with the limitations of the
mass spectrum, lying in the basis of the a geometric approach to
the development of the modified QFT with the Maximal Mass, leads
to the appearance of non-Hermitian Hamiltonians in its fermion
sector (see for example \cite{Max},\cite{KMRS},\cite{Rod1}). This
consequence can be interpreting \textbf{as a necessary condition }
of the finiteness of the mass spectrum (\cite{Markov}).

Thus, it is shown that the \textbf{necessary and a sufficient
conditions} of the limitation of the physical mass of  particles
(\ref{M}) in fermion sector are the using of the non-Hermitian
 $ \cal PT$ - symmetric quantum models with a $\gamma_5$
- massive contribution.

\section{Free fermion models with $\gamma_5$-massive
contributions and  the areas of unbroken $\cal PT$ symmetry}

 Conditions
(\ref{012 }), (\ref{e210})  are executed automatically, if one
enter the following parametrization: \be\label{alpha00} m_1 = m
\cosh(\alpha);\;\;\;\;\ m_2= m \sinh(\alpha). \ee Moreover, from
(\ref{M}) and (\ref{alpha00}) we can express the values of
$m,\,\,m_1,\,\,m_2$ with parameter $\alpha$.

\begin{figure}[h]
\vspace{-0.2cm} \centering
\includegraphics[angle=0, scale=0.5]{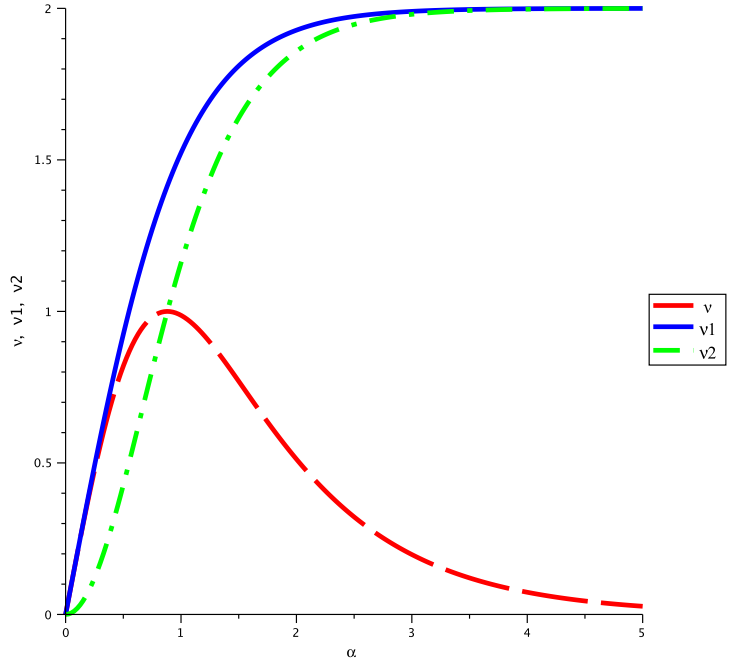}
\caption{Dependence of $\nu=m/m_{max}, \nu_1=m_1/m_{max},
\nu_2=m_2/m_{max} $ from the parameter $\alpha $}
\vspace{-0.1cm}\label{Fig.1}
\end{figure}

 Fig. 1 shows the dependence of the relative values
$\nu=m/m_{max}$, $\nu_1=m_1/m_{max}$ and $\nu_2=m_2/m_{max}$ from
the parameter  $\alpha$. The values of the parameters
$\nu1,\,\nu2$ are the following $\nu\leq\nu_1\leq 2$,
$0\leq\nu_2\leq 2$.  A particle mass $\nu$ may vary in a wide
range $0\leq m\leq m_{max}$. When $\alpha_0 = 0.881 $ it reaches
its maximum, which corresponds to the maximon ($m=m_{max} $).

Using (\ref{M}) and (\ref{alpha00}), you can also get a
\be\label{tan1} \tanh(\alpha) = \sqrt{\frac{1
\pm\sqrt{1-\nu^2}}{2}}. \ee Two of the root sign in (\ref {tan1})
are interpreted as two branches of the values of
$\nu_1(\tilde\nu_1)$ and $\nu_2(\tilde\nu_2)$, which are
multi-valued functions $\nu$. Thus, we have

\be\label{m1} \nu_1(\tilde{\nu_1}) = \sqrt{2} \sqrt{1
\mp\sqrt{1-\nu^2}};
 \ee

\be \label{m2}\nu_2(\tilde{\nu_2}) = \left( 1
\mp\sqrt{1-\nu^2}\right). \ee

It is easy to see, that between these earlier symbols for the
masses of (\ref{dm1}),(\ref{dm2}) and obtained here the values of
(\ref{m1}), (\ref{m2}), after identification of the limiting
masses $M$ and $m_{max}$, there are simple correlations:

\be m_1=m_{max}\nu_1;\,\,\, m_2=m_{max}\nu_2;\ee

\be \tilde{m_1}=m_{max}\tilde{\nu_1}=m_{max}\nu_3=m_3; \,\,\,
\tilde{m_2}=m_{max}\tilde{\nu_2}=m_{max}\nu_4 = m_4.\ee

Fig. 2 demonstrates the dependence of the parameters
$\nu_1(\nu_3)$, $\nu_2(\nu_4)$ from the variable $\nu$. Thus, the
region of the existence of unbroken  $\cal PT$ symmetry can be
represented in the form $0 \leq\nu\leq 1$. For these values of
$\nu$ parameters $\nu_1$ and $\nu_2$ determine the masses of the
modified Dirac equation with a maximal mass $m_{max}$, describing
the  particles having the actual mass $m\leq m_{max}$. However,
the new Dirac equations nonequivalent, because one of them
describes ordinary particles ($\nu_1,\nu_2$), and the other
corresponds their exotic partners($\tilde{\nu_1},\tilde{\nu_2}$).
The special case of Hermiticity is on the line $\nu=1$
($m=m_{max}$ is the case of the maximon), which is the boundary of
the unbroken $\cal PT$ - symmetry. In this point of the plot we
have $\nu_1=\tilde\nu_1=\nu_3=\sqrt{2}$ and
$\nu_2=\tilde\nu_2=\nu_4=1$.

\begin{figure}[h]
\vspace{-0.2cm} \centering
\includegraphics[angle=0, scale=0.5]{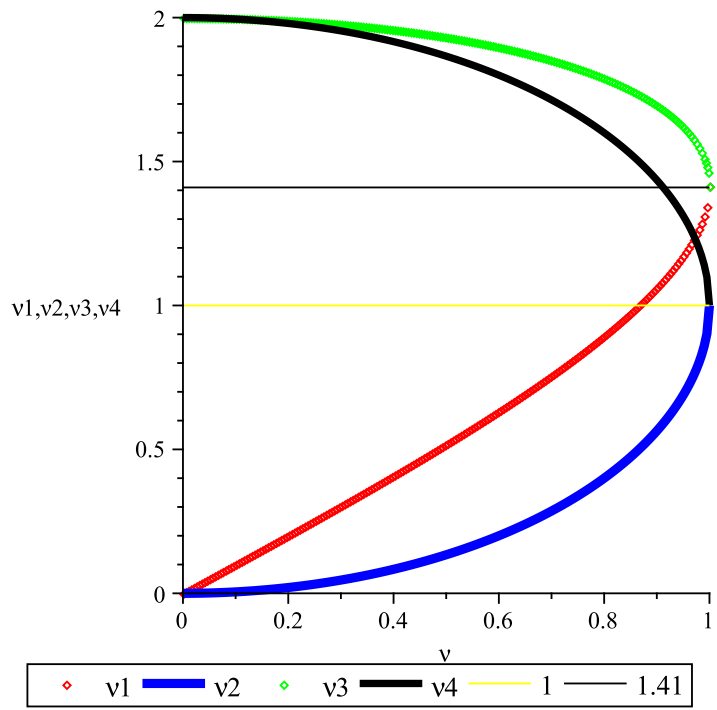}
\caption{The values of parameters   $\nu_1, \nu_2,  \nu_3, \nu_4$
as the function of $\nu$} \vspace{-0.1cm}\label{f4}
\end{figure}

\begin{figure}[h]
\vspace{-0.2cm} \smallskip
\includegraphics[angle=0, scale=0.5]{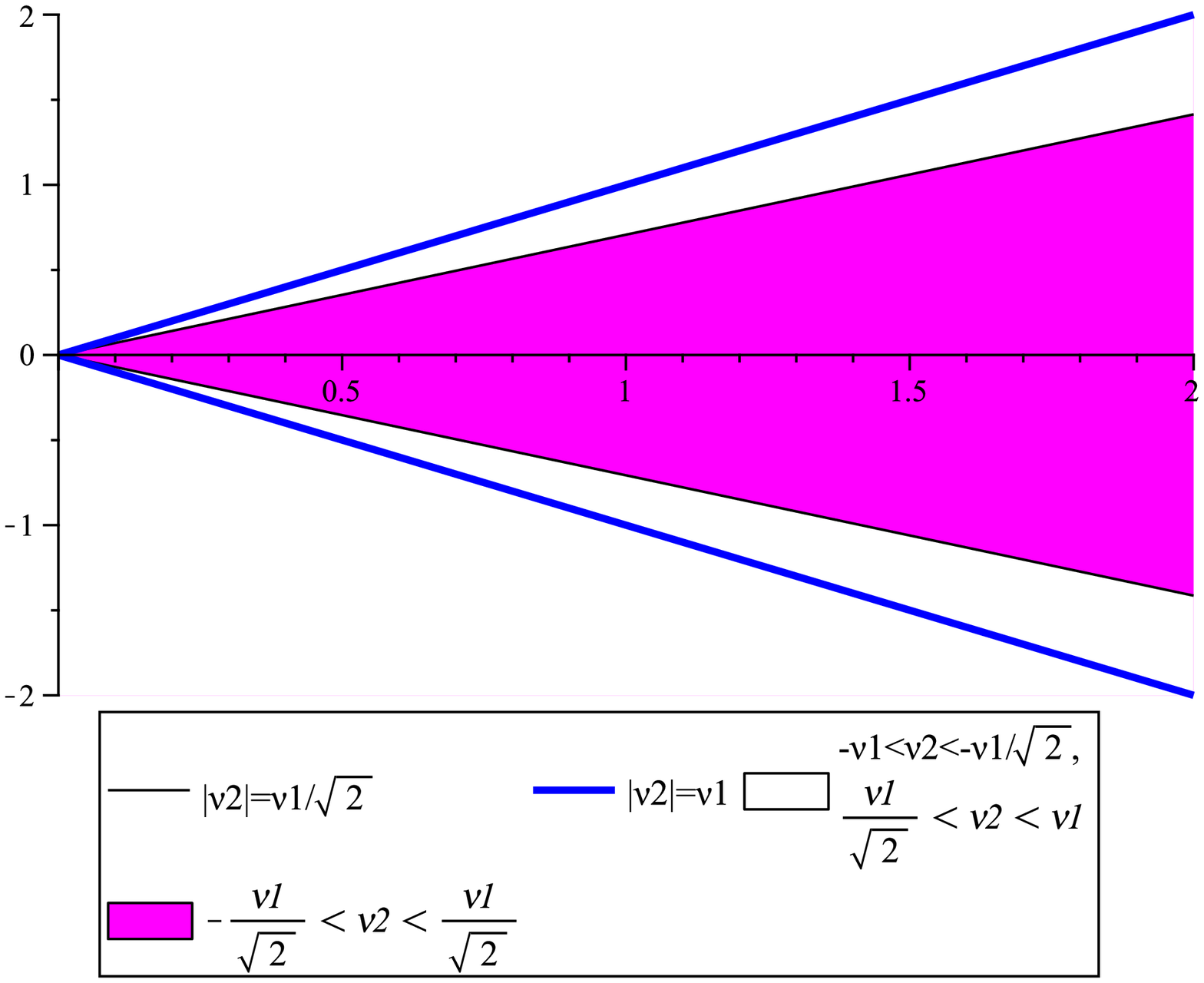}
\caption{The parametric areas of the unbroken $\cal PT$-symmetry
${\nu_1}^2 \geq {\nu_2}^2$ in plane $\nu_1$,$\nu_2$ for the
Hamiltonian (\ref{eps})
%${m_1}^2\geq{m_2}^2$
consists of three specific subregions. Only the shaded area $II.$
meets the ordinary particles, and the bordering with it regions
$I.$ and $III.$ correspond to the description of the exotic
fermions.} \vspace{-0.1cm}\label{f4}
\end{figure}

It is easy to verify that the new values of the mass parameters
$\tilde m_1$, $\tilde m_2$ still satisfy the conditions (\ref{012
}) and (\ref{e210}). We emphasize once again that, from the
formulas (\ref{m1}),(\ref{m2}) in the case of the upper sign it
should be $m_1\rightarrow m$ and $m_2\rightarrow 0$ when $m_{max}
\rightarrow \infty,$ i.e. there is a so-called flat limit,
determined in the frame of the geometric approach
\cite{Max},\cite{KMRS}. However, when one choose a lower sign
(i.e.  for the $\tilde m_1$ and $\tilde m_2$) such a limit is
absent.

 In the frame of the condition ${m_1}^2\geq {m_2}^2$ \quad{} \footnote
 {Note that this inequality was considered in the paper
\cite{ft12} as a single defining expression} at Fig.3 we can see
three specific sectors of unbroken $\cal PT$-symmetry of the
Hamiltonian (\ref{eps}) in the plane $\nu_1, \nu_2$. The plane
$\nu_1,\nu_2$ may be divided  by the three groups of the
inequalities:

$$I.\,\,\,\,\,\,\,\,\,\,\,\,\,\,\,\,\,\,\, \nu_1/\sqrt{2} \leq \nu_2 \leq \nu_1,$$
$$II.\,\,\,\,-\nu_1/\sqrt{2}\leq \nu_2\leq \nu_1/\sqrt{2},$$
$$III. \,\,\,\,\,\,\,\, -\nu_1 \leq \nu_2\leq-\nu_1/\sqrt{2},$$

Only the area $II.$ corresponds to the description of ordinary
particles, then $I.$ and $III.$ agree with the description of some
as yet unknown particles. This conclusion is not trivial, because
in contrast to the geometric approach, where the emergence of new
unusual properties of particles associated with the presence in
the theory a new degree of freedom (sign of the fifth component of
the momentum $p_5$ \cite{KMRS}), in the case of a simple extension
of the free Dirac equation  due to the additional $\gamma_5$-mass
term, the satisfactory explanation is not there yet.

\section{Conclusion}

Starting with the researches, presented in the previous sections,
we have shown that Dirac Hamiltonian of a particle with $\gamma_5$
- dependent mass term is non-Hermitian, and has the unbroken $\cal
PT$ - symmetry in the area ${m_1}^2\geq {m_2}^2,$ which has three
of a subregion. Indeed with the help of the algebraic
transformations we obtain a number of the consequence of the
relation (\ref{012 }). In particular there is the restriction of
the particle mass in this model $m\leq m_{max}$, were
$m_{max}={m_1}^2/2 m_2$. Outside of this area the $\cal PT$ -
symmetry of the modified Dirac Hamiltonians is broken.

 In addition, we have shown
that the introduction of the postulate about the limitations of
the mass spectrum, lying in the basis of the a geometric approach
to the development of the modified QFT (see, for example
\cite{Max},\cite{KMRS}), leads to the appearance of non-Hermitian
$\cal PT $-symmetric Hamiltonians in the fermion sector of the
model with the Maximal Mass. Thus, it is shown that  using of
non-Hermitian $\cal PT$-symmetric quantum theory with $\gamma_5$
mass term may be considered as \textbf{ necessary and sufficient
conditions} the appearance of the limitation of the mass particle
(\ref{M}) in a fermion sector of the model.

In particular, this applies to the modified Dirac equation in
which produced the substitution $m\rightarrow m_1+ \gamma_5 m_2 $.
Into force of the ambiguity of the definition of parameters $m_1,
m_2$ the inequality $m_1\geq m_2$ describes a particle of two
types. In the first case, it is about ordinary particles, and when
$m_1,m_2\geq 0$ mass parameters  are limited by the terms

\be \label{01}     0\leq m_2 \leq m_1/\sqrt{2} . \ee

In the second area we are dealing with so-called «exotic partners»
of ordinary particles, for which is still accomplished
(\ref{e210}), but one can write

\be \label{02} m_1/\sqrt{2} \leq m_2 \leq m_1. \ee

Intriguing difference between particles of the second type from
traditional fermions is that they are described by the other
modified Dirac equations. So, if in the first case(\ref{01}), the
equation of motion there has a limit transition when $m_{max}
\rightarrow \infty$ that leads to the standard Dirac equation,
however in the inequality (\ref{02}) such a limit is not there.

Thus, it is proved that the main progress, obtained by us the in
the algebraic way of the construction of the fermion model with
$\gamma_5$ - dependent mass term applies to   the limitations of
the mass spectrum. Furthermore, the possibility of describing of
the exotic particles are turned out essentially the same as in the
model with a maximal mass, which was investigated by
V.G.Kadyshevsky with colleagues \cite{Kad1} - \cite{Rod} on the
basis of geometrical approach. It is also shown that the
transition point at the scale
 the masses from the ordinary particles to the exotic this is  mass of
 the maximon.

On the basis of (\ref{M}) it also has been shown that the
parameters $ m_1$ and $ m_2$ have the auxiliary nature. This fact
is easily proved by means of the comparison of  the ordinary and
exotic fermion fields. Thus, it is the important conclusion   that
the description of exotic fields may be considered  with the help
of the algebraic approach and  is not the prerogative of the
geometric formalism. Note that the polarization properties of the
exotic fermion fields fundamentally differ from  the standard
fields that with taking into consideration of interactions  may be
of interest in the future  researches.

{\bf Acknowledgment:} We are grateful to Prof. V.G.Kadyshevsky for
fruitful and highly useful discussions.

 \end{document}